\documentclass[12pt] {article}
\def\be{\begin{equation}}
\def\ee{\end{equation}}
\def\bea{\begin{eqnarray}}
\def\eea{\end{eqnarray}}
\def\pd{\partial}
\begin{document}
\title{Lagrange Brackets and $U(1)$ fields}
\author{David B. Fairlie$\footnote{e-mail: david.fairlie@durham.ac.uk}$\\
Department of Mathematical Sciences,\\
         Science Laboratories,\\
         University of Durham,\\
         Durham, DH1 3LE, England}
\date{May 1st, 2000}
\maketitle
\begin{abstract}
The idea of a companion Lagrangian associated with  $p$-Branes is extended to include the presence of  $U(1)$ fields. The Brane Lagrangians are constructed with $F_{ij}$ represented in terms of Lagrange Brackets, which make manifest the reparametrisation invariance of the theory; these are replaced by Poisson Brackets in the companion Lagrangian, which is now covariant under field redefinition. The ensuing Lagrangians possess a similar formal structure to those in the absence of an anti-symmetric field tensor.
\end{abstract}

\section{Introduction} 
Recently we took up an idea for the association of a field theory with 
Strings and Branes  \cite{bf1,bf2}, which originated with Hosotani \cite{hos1,hos2}. Our motivation was an attempt to answer the question
of to how to construct for the String and Brane Lagrangians, a field theory analogous to the correspondence between the classical particle Lagrangian
and the Klein Gordon Lagrangian.  Relying upon the idea that since the String/Brane Lagrangian is reparametrisation invariant the natural property to be expected of the corresponding field theory Lagrangian, which we call the {\it
companion Lagrangian} is that it should be covariant under general field redefinition: a $p$-Brane with a $p+1$ dimensional world volume embedded in a $d$-dimensional spacetime with Lagrangian
\be
\sqrt{\det\left|\frac{\pd X^\mu}{\pd x_i}\frac{\pd X^\mu}{\pd x_j}\right|}\label{bornbrane} 
\ee
should also be  related to a Lagrangian for $p+1$ fields $\phi^i$
with Lagrangian
\be
{\cal L}=\sqrt{\det\left|\frac{\pd \phi^i}{\pd x_\mu}\frac{\pd \phi^j}{\pd x_\mu}\right|}.\label{inersebrane}
\ee
A different association was made by Morris  \cite {morris1}, who conceptualised the String as the intersection of $d-2$ hypersurfaces $\phi^j$=constant. The
ensuing Lagrangian takes the same form as above, with  but for $d-2$ fields and corresponds to the interchange of dependent and independent variables in (\ref{bornbrane}). The equations of motion are then simply transforms of those for (\ref{bornbrane}). This suggestion has, in view of our motivation, the undesirable feature that the number of fields is dependent upon the dimensions of the total space and is therefore not the natural extension of the Klein Gordon Field Theory. In our second paper \cite{bf2} we derived the Lagrangian  (\ref{inersebrane}) from the the Hamilton-Jacobi equation for Strings and Branes \cite{nambu,kastrup,spall}
together with some additional constraints. This provides further justification for the companion Lagrangian.  We obtained an additional refinement;
the dimensions of the space upon which the Lagrangian is defined are reduced to
$d-1$. This we proved for the cases of one and two fields; a general proof has subsequently been given by L. Baker \cite{linda} 
  
\section{Reparametrisation Invariance, Covariance}
Kastrup in his review, \cite{kastrup2} quotes literature going back to Caratheodory on the properties of Lagrangians with reparametrisation invariance.
These are characterized by the property
\be
\sum_a\phi^a_j\frac{\pd {\cal L}}{\pd\phi^a_k}\,=\,\delta_{jk}{\cal L}.\label{rep}
\ee
The corresponding characterization of covariant Lagrangians  is that they should satisfy the following conditions;
\be
\sum_j\phi^a_j\frac{\pd {\cal L}}{\pd\phi^b_j}\,=\,\delta_{ab}{\cal L}\label{cov}.
\ee
These latter constraints remain invariant under field redefinition, and define Lagrangian densities  homogeneous of weight one.
The sum is over lower indices now instead of upper indices. Caratheodory has shown that such reparametrisation invariant Lagrangians must be built out of the Jacobians of the fields, as is also the case for those of weight one. The determinants in both (\ref{bornbrane}) and (\ref{inersebrane}) may be expressed as the sums of squares of Jacobians \cite{dbf}.It is easy to verify that these Lagrangians satisfy (\ref{rep}) and (\ref{cov}) respectively. If the Lagrangian is both reparametrisation invariant and covariant, then $\phi^a_j$ is of the form of a square invertible matrix and the above relations imply that ${\cal L}$ is 
the Jacobian of the $\phi^a$ with respect to the co-ordinates $x_j$ and is therefore a divergence, making the theory completely topological. 
\section{Brane Lagrangians and their Companions}
The next objective is to extend the ideas of \cite{bf2} to include gauge fields. This is not so trivial as it might seem as the properties of reparametrisation and covariance under field redefinition respectively must be respected.
The Brane Lagrangian is now
\be 
{\cal L}\,=\,\sqrt{\det{| g_{ij}+F_{ij} |}}\,=\, \sqrt{\det\left| \frac{\partial X^\mu}{\partial x_i}\frac{\partial X_\mu}{\partial x_j}\,+\,\left( \frac{\partial A_j}{\partial x_i}\,-\,\frac{\partial A_i}{\partial x_j}\right)\right|}\ .\label{born}
\ee
Let us consider instead
\be 
{\cal L}\,=\,\sqrt{\det{| g_{ij}+F_{ij} |}}\,=\, \sqrt{\det\left| \frac{\partial X^\mu}{\partial x_i}\frac{\partial X_\mu}{\partial x_j}\,+\,\left( \frac{\pd p}{\partial x_i}\frac{\pd q}{\partial x_j}\,-\,\frac{\pd p}{\partial x_j}\frac{\pd q}{\partial x_i}\right)\right|}\ .\label{born2}
\ee
The field $F_{ij}$ has been replaced by a Lagrange Bracket. It  still describes a $U(1)$ theory, but one in which $\epsilon_{ijkl}F_{ij}F_{kl}\,=\,0$ for all selections of four out of the $p+1$ indices $i,j\dots$ etc.
The property that the expression under the square root is the sum of squares still holds, as it depends only upon the antisymmetry of $F_{ij}$. 
Here 
\be
F_{ij}\,=\,  \frac{\pd p}{\partial x_i}\frac{\pd q}{\partial x_j}\,-\,\frac{\pd p}{\partial x_j}\frac{\pd q}{\pd x_i}\, =\, [D_i,\ D_j],\label{com}
\ee
where the covariant derivative $D_i$ is defined by
$$
D_i = \frac{\pd }{\pd x_i} + p\frac{\pd q}{\pd x_i}.
$$
This is for a 2-dimensional phase space; by taking a multidimensional phase space, using
$$
D_i = \frac{\pd }{\pd x_i} + \sum_\alpha p^\alpha\frac{\pd q_\alpha}{\pd x_i}.
$$
any U(1) gauge field $A_i$ can be expressed as the  connection in the formula above and the restriction upon $F\wedge F$ is relaxed. (In higher dimensions,
the number of phase space variables present dictates the order of non vanishing
products of the form $F\wedge F,\ F\wedge F\wedge F$  etc which can appear). 
Note that it is evident that (\ref{born2}) is reparametrisation covariant,
i.e. transforms as the Jacobian of the transformation of co-ordinates.
Note also that (\ref{born2}) possesses the same formal structure as in the absence of the gauge field as it may be written as
\be
{\cal L}\,=\, \sqrt{\det\left| \frac{\partial X^\alpha}{\partial x_i}g_{\alpha\beta}\frac{\partial X^\beta}{\partial x_j}\right|}\label{born3}
\ee
where the metric $g_{\alpha\beta}$ is a direct sum of the Lorentz (Euclidean)
metric and a symplectic one;
$$
  g_{\alpha\beta}\,=\, \eta_{\mu\nu}\oplus\epsilon_{ab}.
$$
An analogous metric is easily constructed in the case of a multidimensional phase space.
What would be the corresponding companion Lagrangian, in the sense of
\cite{bf1,bf2}?
The property any such Lagrangian should fulfil is that it is covariant; i.e. transforms with the Jacobian of the fields.
One suggestion which is rather different from the possibilities discussed in 
\cite{bf1} is to take 
\be 
{\cal L}\,=\, \sqrt{\det\left| \frac{\partial \phi^i}{\partial x_\mu}\frac{\partial \phi^j}{\partial x_\mu}\,+\,\left( \frac{\pd \phi^i}{\partial p}\frac{\pd \phi^j}{\partial q}\,-\,\frac{\pd \phi^j}{\partial p}\frac{\pd\phi^i}{\partial q}\right)\right|}\ .\label{comp}
\ee
This is manifestly covariant and the corresponding  antisymmetric field
$$
F^{ij}\,=\,  \frac{\pd \phi^i}{\partial p}\frac{\pd \phi^j}{\partial q}\,-\,\frac{\pd \phi^j}{\partial p}\frac{\pd\phi^i}{\partial q}
$$
is here expressed as a Poisson Bracket, and is the commutator term in an $SU(\infty)$ gauge theory \cite{ffz}. Recall that Poisson and Lagrange brackets are inverses of each other, when the dimensions of phase space match those of the co-ordinates. The fact that this is an infinite gauge group might allay the possible objections to the concept of a companion field theory, on the grounds that String Field theorists want to have a field which is a functional of the the string length, in order to match the number of degrees of freedom. Obviously the replacement of $p,\ q$ by matrices permits the construction of a non-Abelian field in (\ref{com}), but it is not clear how to incorporate this into a Lagrangian. The resolution of this problem might afford another approach to 
the construction of Born-Infeld Lagrangians with non-Abelian gauge fields\cite{tseyt}. This ansatz for the gauge fields is also reminiscent of the ADHM construction for instantons where the gauge fields are represented in terms of rectangular matrices. \cite{adhm}

\section{Lagrangian determinant as a quadratic}
First let me remind the reader of the way in which the 4-dimensional case of the Dirac-Born-Infeld Lagrangian may be expressed as a sum of squares from \cite{dbf}
\bea 
L^2&=&\frac{1}{4!}\sum_{\mu\nu\rho\sigma} \left(\epsilon_{ijkl}\frac{\partial X^\mu}{\partial x_i}  \frac{\partial X^\nu}{\partial x_j}\frac{\partial X^\rho}{\partial x_k}  \frac{\partial X^\sigma}{\partial x_l}\right)^2\nonumber\\
&+& \frac{1}{2!}\sum_{\mu\nu} \left(\frac{1}{2}\epsilon_{ijkl}\frac{\partial X^\mu}{\partial x_i}  \frac{\partial X^\nu}{\partial x_j}F_{kl}\right)^2
\,+\, \left(\frac{1}{4}\epsilon_{ijkl}F_{ij}F_{kl}\right)^2\label{quadrat}
\eea
 This expression shows the pattern; the individual terms may be expressed as the contraction of an epsilon symbol with factors of the form $\frac{\partial X^\mu}{\partial x_i}$ ,or else   $F_{ij}$. The case of general dimension is similar. Now consider the case where $F_{ij}$ is represented as a Lagrange Bracket with a 2-dimensional phase space, i.e. a single $p,\ q$. 
The above expression (\ref{quadrat}) holds, but all terms with more than one $F_{ij}$ vanishes, so we have either none or one! This statement can be recast as follows;
The term under the square root may be decomposed  into a sum of squares as
$$\frac{1}{n!}\sum_{\rm permutations}\left(\epsilon_{i_1,i_2,\dots i_n}\frac{\pd X^{\mu_1}}{\pd x_{i_1}}\frac{\pd X^{\mu_2}}{\pd x_{i_2}}\cdots\frac{\pd X^{\mu_n}}{\pd x_{i_d}}\right)^2
$$
for the Brane, where for two of the indices $\mu_1,\ \mu_2$, say $X^{\mu_1}$ may be identified with $p$ and  $X^{\mu_2}$ with $q$ but if $p$ appears, so must $q$. The corresponding companion determinant may be expressed , mutatis mutandis, as
$$\frac{1}{n!}\sum_{\rm permutations}\left(\epsilon_{i_1,i_2,\dots i_n}\frac{\pd\phi^{i_1}}{\pd x_{\mu_1}}\frac{\pd \phi^{i_2}}{\pd x_{\mu_2}}\cdots\frac{\pd\phi^{i_n}}{\pd x_{\mu_d}}\right)^2
$$
where for two of the indices $\mu_1,\ \mu_2$, say, $x_{\mu_1}$ may be identified with $p$ and  $x_{\mu_2}$ with $q$. This means in effect, that the case with a gauge field is in fact equivalent to the case without, except for an increase in dimensions of the larger space and the restriction  that the terms with  co-ordinates $p$ must also have the corresponding $q$ present and vice-versa. 
\section{Conclusions}
The idea of associating a companion Lagrangian with Strings and Branes is further strengthened by the observation that the association also extends to the incorporation of $U(1)$ fields, via the representation of such fields in terms of Lagrange brackets. The corresponding fields which appear in the companion Lagrangian are represented in terms of Poisson Brackets, the large $N$ limit of 
$SU(N)$ commutators. The representation of gauge fields by Lagrange brackets
is a subject for further exploration. Just as the Poisson bracket admits a Moyal deformation,  the Lagrange Bracket will admit a similar associative deformation,
 but in both cases either covariance or reparametrisation invariance are apparently lost. A remarkable feature of the formalism developed is the fact that the Lagrangians presented here have such a similar structure to those in the absence of the  $U(1)$ fields.
The augmentation of
the target space by the phase space in the Brane Lagrangian (or of the base space in the companion Lagrangian) might be a fruitful avenue for further speculation in terms of the concepts of dimensional reduction.

\end{document}